%% file: main.tex
\documentclass[11pt,a4paper]{article}
\usepackage[utf8]{inputenc}
\usepackage[T1]{fontenc}

\usepackage{xcolor} 
\usepackage{indentfirst}
\usepackage{dsfont}    
\usepackage{tcolorbox} 
\usepackage{appendix} 
\usepackage{booktabs}
\usepackage{amsmath, amssymb} 
\usepackage{authblk}          
\usepackage[colorlinks=true, allcolors=blue]{hyperref}
\usepackage{graphicx}
\usepackage{subcaption}
\newcommand{\emailAdd}[1]{{\ttfamily\href{mailto:#1}{#1}}}

\definecolor{lightgreen}{RGB}{144,238,144}
\definecolor{lightblue}{RGB}{173,216,230}

\title{\boldmath $E$ and $J$ type $\mathcal{N}=(0,2)$ disordered models and higher-spin symmetry}

\author[$\heartsuit$]{Liang Wang\thanks{\emailAdd{wangleonard.021@gmail.com}}}
\author[$\diamondsuit$]{Miao Wang\thanks{\emailAdd{wangmiao21@mails.ucas.ac.cn}}}

\affil[$\diamondsuit$]{ Kavli Institute for Theoretical Sciences (KITS), University of Chinese Academy of Sciences, Beijing 100190, China}
\affil[$\heartsuit$]{ School of Physics and Astronomy, Shanghai Jiao Tong University, Shanghai 200240, China}
\date{}

\begin{document} 

\maketitle 
\flushbottom

\begin{abstract}
  In this work, we investigate the emergence of higher-spin structure in 2d $\mathcal{N}=(0,2)$ disordered models. While previous studies focused on the $J$-type model where the $E$-term in the Fermi multiplet was discarded. We extend the discussion to  $\mathcal{N}=(0,2)$ disordered models with $E$-type potential. In terms of (disordered) $\mathcal{N}=(0,2)$ Landau-Ginzburg theory, we establish a duality between two models. By solving the Schwinger-Dyson equations and the ladder kernel matrix for 4-point functions, we verify that the $E$-type model is dynamically equivalent to the $J$-type model in the IR regime. Furthermore, we demonstrate that the $E$-type model also exhibits emergent higher-spin symmetry in certain limits. Our results reveal a larger region of the moduli space of 2D $\mathcal{N}=(0,2)$ disordered theories and provides insights into the holographic transition from finite to tensionless strings that can be diagnosed by the emergence of higher-spin symmetries.
\end{abstract}

\input{Main/intro}

\input{Main/LG_Dual}

\input{Main/E_type_calculation}

\input{Main/J_type_an_higher_spin}

\input{Main/summary}

\bibliographystyle{JHEP}
\bibliography{ref}
\appendix
\input{Main/appendix}

\end{document}

%% file: Main/intro.tex
\section{Introduction}

The Sachdev-Ye-Kitaev (SYK) model is a prominent disordered system, serving as a rare example of a theory that is simultaneously strongly coupled yet perturbatively solvable. Crucially, the model exhibits an emergent reparameterization symmetry in the IR. 
The spontaneous breaking of this symmetry generates soft modes governed by a Schwarzian action, which establishes a holographic duality with JT gravity on near-AdS$_2$ spacetimes~\cite{Sachdev_1993,Kitaev:2015,Maldacena_2016,maldacena2016conformalsymmetrybreakingdimensional,Rosenhaus_2019}.

While a precise top-down string realization of the SYK model remains under exploration, higher-spin theories can serve as a crucial conceptual bridge. Specifically, higher-spin theories represent the tensionless limit of string theory~\cite{Gaberdiel_2014,Gaberdiel_2016}, whereas the SYK model is believed to act as a holographic dual to string theory with finite tension~\cite{Maldacena_2016}. Furthermore, the observed finite anomalous dimensions suggest that the SYK model can be interpreted as a deformation of vector models that feature a tower of higher-spin operators. A notable example exhibiting such SYK-like characteristics is the Gross-Neveu vector model, which was studied in detail in~\cite{Peng_2017}. In fact, examples of 1d disordered theory that demonstrates an explicit transition between an integrable phase, where a large number of conserved quantities exist similar to the higher-spin theory, and a chaotic phase are constructed explicitly~\cite{Gao:2024lve} .

In higher dimensions, disordered quantum field theories could also exhibit the emergence of higher-spin symmetries on the boundary of the theories' moduli space~\cite{Peng_2018, Ahn:2018sgn,Chang:2021fmd,Chang:2021wbx}. These examples all have certain number of supersymmetries, which is common in high dimensional disordered models. 
Compared with one-dimensional supersymmetric SYK models~\cite{Fu:2016vas,Peng:2016mxj, Li:2017hdt, Peng:2017spg, Peng_2017,Bulycheva:2018qcp, Peng:2020euz,Heydeman:2022lse,Heydeman:2024ohc}, higher dimensional covariant supersymmetric SYK model~\cite{Murugan:2017eto,Peng_2018, Chang:2021fmd,Chang:2021wbx} exhibits even more intriguing characteristics. In two dimensions, the factorization of the isometry into left- and right-moving sectors allows for a clearer exploration of the moduli space~\footnote{This property is also crucial in the construction of other 2d SYK like models without supersymmetry, see e.g.~\cite{Berkooz:2016cvq,Turiaci:2017zwd}.}, while simultaneously manifesting distinct higher-spin properties in the $\mathcal{N}=(0,2)$ SYK-like disordered models proposed in~\cite{Peng_2018}. Higher-spin symmetry with $\mathcal{N} = (0,2)$ supersymmetry has also been observed in 3d disordered models in~\cite{Chang:2021fmd,Chang:2021wbx}.

Higher-spin gravity theories in 2+1 d are conjectured to be dual to the 't-Hooft limit of 2d minimal model CFTs~\cite{Gaberdiel_2011,Gaberdiel_2014,Candu_2013,Creutzig_2012}. While extensive studies have addressed such dualities involving both left- and right-moving sectors, the $\mathcal{N}=(0,2)$ HS AdS$_3$ gravity sector was only recently investigated in~\cite{Cao:2025dja}. Complementing this gravity-side analysis, our work focuses on the dual CFT description, specifically employing a disordered model in the conformal regime. As demonstrated in~\cite{Peng_2018}, the $\mathcal{N}=(0,2)$ disordered model exhibits emergent higher-spin symmetry in the absence of the Fermi multiplet $E$-field—a configuration we refer to as the $J$-type model.

In this work, we refine the discussion by focusing on the pure $E$-model configuration where $E \neq 0$ and $J = 0$. A primary motivation for this study is that a rigorous derivation of the $E$-model directly from Feynman diagrams provides the necessary mathematical foundation for exploring more complex mixed $E-J$ models in the future. We demonstrate that in the IR regime, this model exhibits higher-spin properties identical to those of the pure $J$-model, thereby effectively broadening the moduli space of $\mathcal{N} = (0,2)$ higher-spin theories in disordered systems.
	
To substantiate this, we organize the paper as follows. In Section 2, we interpret the duality between the two models using the $(0,2)$ Landau-Ginzburg theory. In Section 3, we derive the Schwinger-Dyson equations. In Section 4, we compute the integral kernels explicitly from the fundamental Feynman diagrams; we show that despite the fundamentally different topological connections compared to the $J$-model, the exact matching of the kernel eigenvalues proves the reliability of our calculations and the expected duality. In Section 5, we confirm the higher-spin nature of the $E$-type model through concrete numerical verification. Finally, we conclude our results in Section 6.

%% file: Main/LG_Dual.tex
\section{The $(0,2)$ Landau-Ginzburg Model} \label{LGDual}

The action $S_{\Lambda}+S_{\Phi}+S_{J}$, as presented in~\cite{Peng_2018}, describes a specific $(0,2)$ Landau-Ginzburg model characterized by $J(\Phi)=J_{ia_1\cdots a_q}\Phi_{a_1}\cdots\Phi_{a_q}$ and $E=0$. Although general $(0,2)$ theories may involve Chiral ($\Phi$), Fermi ($\Lambda$), Vector ($V$), and Gauge ($\Upsilon$) multiplets~\cite{Witten_1993}, our study concentrates on models containing only $\Phi$ and $\Lambda$ interacting through $J(\Phi)$. Such models define a rich landscape of symmetry and topology~\cite{Melnikov_2008}. Specifically, this section provides an explicit analysis of the $E \leftrightarrow J$ symmetry at the action level.

\subsection{Setup Superfields}

We adopt the notations and conventions as in \cite{Peng_2018}. Throughout this work, we operate in two-dimensional Euclidean space parameterized by the complex coordinates $z = x_0 + ix_1$ and $\bar{z} = x_0 - ix_1$. The component expansions for the chiral ($\Phi$) and Fermi ($\Lambda$) superfields are then expressed as follows,
\begin{align*}
	\Phi^i &= \phi^i+ \sqrt{2} \theta^+ \psi^i +2 \theta^+ \bar{\theta}^+ \partial_z \phi^i, \\
	\bar{\Phi}^i &= \bar{\phi}^i - \sqrt{2} \bar{\theta}^+ \bar{\psi}^i - 2\theta^+ \bar{\theta}^+ \partial_z \bar{\phi}^i, \\
	\Lambda^a &= \lambda^a - \sqrt{2} \theta^+ G^a +2 \theta^+ \bar{\theta}^+ \partial_z \lambda^a - \sqrt{2} \bar{\theta}^+ E^a(\Phi), \\
	\bar{\Lambda}^a &= \bar{\lambda}^a - \sqrt{2} \bar{\theta}^+ \bar{G}^a -2 \theta^+ \bar{\theta}^+ \partial_z \bar{\lambda}^a - \sqrt{2} \theta^+ \bar{E}^a(\bar{\Phi}).
\end{align*}

The field content consists of $N$ chiral supermultiplets and $M$ Fermi supermultiplets, labeled by the indices $i=1,\dots,N$ and $a=1,\dots,M$, respectively. We define the ratio $\mu \equiv M/N$, which serves as the higher-spin parameter in later discussions. These fields, along with their Hermitian conjugates, satisfy the standard supersymmetry constraints. By introducing the $\mathcal{N}=(0,2)$ superspace covariant derivatives $D_+ = \frac{\partial}{\partial \theta^+} +2\bar{\theta}^+ \partial_z$ and $\bar{D}_+ = -\frac{\partial}{\partial \bar{\theta}^+} + 2 \theta^+ \partial_z$, the constraints are explicitly given by:

\begin{align}
\bar{D}_+ \Phi &= 0, \\
\bar{D}_+ \Lambda &= \sqrt{2}E, \\
\bar{D}_+ E &= 0, \\
E \cdot J &= 0\label{constraintsEJ}
\end{align}

Here are some comments on $E$ field. It is worth noting that $E = 0, J \neq 0$ and $E \neq 0, J = 0$ are the two most intuitive ways to satisfy the constraint $E \cdot J = 0$. However, once we turn on $E$ field, the constraints become inhomogeneous and complicate the theory. Aside from that, it would be convenient to work in the component formalism for $E(\Phi)$ field so we do the following expansion (similar rules also apply to $J(\Phi)$),
\begin{equation*}
E^a(\Phi) = E^a(\phi) + \sqrt{2} \theta^+ E^a_{,j} \psi^j +2\theta^+ \bar{\theta}^+ E^a_{,j} \partial_z \phi^j.
\end{equation*}

Here, $E^a_{,j} \equiv \partial E^a / \partial \phi^j$ and indices $i, j$ run over the Chiral multiplets, and $a$ runs over the Fermi multiplets. We obtain Fermi multiplets in component form,
\begin{align*}
\Lambda^a &= \lambda^a - \sqrt{2} \theta^+ G^a +2\theta^+ \bar{\theta}^+ \partial_z \lambda^a - \sqrt{2} \bar{\theta}^+ E^a(\phi) + 2 \theta^+ \bar{\theta}^+ E^a_{,j} \psi^j, \\
\bar{\Lambda}^a &= \bar{\lambda}^a - \sqrt{2} \bar{\theta}^+ \bar{G}^a - 2\theta^+ \bar{\theta}^+ \partial_z \bar{\lambda}^a - \sqrt{2} \theta^+ \bar{E}^a(\bar{\phi}) + 2 \theta^+ \bar{\theta}^+ \bar{E}^a_{,j} \bar{\psi}^j.
\end{align*}

\subsection{Symmetry in the $(0,2)$ Landau Ginzburg Action}

Consistent with the result in~\cite{DISTLER1994213,Peng_2018,adams200402duality},we expand action $S_{\Phi} + S_{\Lambda} + S_{J}$ as follows,
\begin{align*}
S_{\Phi} &\equiv - \int d^2z \int d\theta^{+}d\bar{\theta}^{+} \bar{\Phi} \partial_{\bar{z}} \Phi \nonumber \\
&= \int d^2z \left( 4 \bar{\phi} \partial^{2} \phi - 2 \bar{\psi} \partial \psi \right), \\
S_{\Lambda} &\equiv \frac{1}{2} \int d^2z \, d\theta^{+}d\bar{\theta}^{+} \bar{\Lambda} \Lambda \nonumber \\
&= \int d^2z \left(- 2\bar{\lambda} \partial_{z} \lambda + \bar{G}G - \bar{E}E - \bar{\lambda}_i E^i_{,j}\psi^j - \bar{E}^i_{,j}\bar{\psi}^j \lambda_i\right),\\
S_J &\equiv -\int d^2z\, d\theta^+ \Lambda^i J_i(\Phi)|_{\bar{\theta}^+=0} + \text{h.c.}\\
&=\sqrt{2} \int d^2z \left(\lambda^i J_{i,j} \psi^j + G^i J_i + \bar{\psi}^j \bar{J}_{i,j} \bar{\lambda}^i + \bar{J}_i \bar{G}^i\right).
\end{align*}

It is worth noting that, up to this point, no specific forms for $E$ and $J$ have been assumed. The absence of coupling terms between them in $S_J$ is guaranteed by constraint in Eq.~\eqref{constraintsEJ}. Consequently, the final complete Lagrangian $\mathcal{L}$ is given by,

\begin{equation} \label{totalLag}
\begin{split}
\mathcal{L} &= \left( 4 \bar{\phi} \partial^{2} \phi - 2 \bar{\psi} \partial \psi \right) \\
&\quad + \left(- 2\bar{\lambda} \partial_{z} \lambda + \bar{G}G - \bar{E}E - \bar{\lambda}_i E^i_{,j}\psi^j - \bar{E}^i_{,j}\bar{\psi}^j \lambda_i \right) \\
&\quad + \sqrt{2} \left(\lambda^i J_{i,j} \psi^j + G^i J_i + \bar{\psi}^j \bar{J}_{i,j} \bar{\lambda}^i + \bar{J}_i \bar{G}^i \right).
\end{split}
\end{equation}

We observe that the $G$ field lacks kinetic terms, functioning as an auxiliary field. Integrating it out by solving the equations of motion yields $G^i = -\sqrt{2} \bar{J}_i$ and $\bar{G}^i = -\sqrt{2} J_i$. Substituting these expressions back into the action, the effective Lagrangian becomes,
\begin{align*}
\mathcal{L} &= \mathcal{L}_{\text{kin}(\phi,\psi,\lambda)} \\
&- 2 \bar{J} J - \bar{\psi}^j (-\sqrt{2} \bar{J}^i_{,j}) \bar{\lambda}_i - \lambda_i (-\sqrt{2} J^i_{,j}) \psi^j \\
&- \bar{E} E - \bar{\psi}^j \bar{E}^i_{,j} \lambda_i - \bar{\lambda}_i E^i_{,j} \psi^j.
\end{align*}

Comparing the $J$-terms and $E$-terms, a clear symmetry transformation emerges. This duality is explicitly realized through the identification $E \Leftrightarrow -\sqrt{2}J$ and the fermionic exchange $\lambda \leftrightarrow \bar{\lambda}$.

%% file: Main/E_type_calculation.tex
\section{Calculations in the $E$-type Model}\label{sykcal}

Although we have identified a structural correspondence between the
models, the specific conditions for the mapping $\lambda\leftrightarrow \bar \lambda$, as well as potential
discrepancies in coefficients, need careful examination to determine whether
they lead to physically distinct predictions. In this section, we verify the con
sistency of the Schwinger-Dyson (SD) equations in the infrared (IR) regime.
\subsection{Model Settings and Action}

For the $E$-type model, we set $J=0$ and ask $E(\Phi)$ to carry the quenched disorder $E_a(\Phi_i) = J_{ai_1i_2,\ldots,i_q} \Phi_{i_1} \ldots \Phi_{i_q}$. The summation over repeated indices is orderless, and the Gaussian random variables $J_{ia_1\cdots a_q}$ satisfy the following statistics,
\begin{align}
    \langle J_{ia_1\ldots a_q}\rangle &= 0, \\
    \langle J_{ia_1\ldots a_q} \bar{J}_{ia_1\ldots a_q} \rangle &= \frac{(q-1)!}{N^q} J^2.
\end{align}

Referring to Eq. \eqref{totalLag}, the Lagrangian is given by,
\begin{equation*} 
\begin{split}
\mathcal{L} &= \mathcal{L}_{\text{kin}} - \bar{E}E - \bar{\lambda}_i E^i_{,j}\psi^j - \bar{E}^i_{,j}\bar{\psi}^j \lambda_i.
\end{split}
\end{equation*}

To integrate the disorder, we must evaluate Gaussian integrals over $J_{ai_1i_2,\ldots,i_q}$. However, the explicit $\bar{E}E$ term is inconvenient for these calculations. To address this, we introduce an auxiliary field $B$ to linearize the term via the Hubbard-Stratonovich transformation,
\begin{equation*}
-\bar{E}E \sim \bar{B}B - BE - \bar{B}\bar{E}.
\end{equation*}

Thus, the complete effective action becomes,
\begin{equation}\label{elage}
    \mathcal{L}\cong \, \mathcal{L}_{\text{kin}} + \bar{B}B - BE - \bar{B}\bar{E} - \bar{\psi}^j \bar{E}_{a,j} \lambda^a - \bar{\lambda}^a E_{a,j} \psi^j.
\end{equation}

\subsection{$G\Sigma$ Action and SD Equations}
We now compare the linearized effective action of the E-type model with that of the J-type model presented in \cite{Peng_2018}:
\begin{align*}
E\text{-type model, } & \quad {-J_{a i_1 \cdots i_q} \cdot \left(q \bar{\lambda}^a \psi^{i_1} \phi^{i_2} \cdots \phi^{i_q} +  B^a \phi^{i_1} \phi^{i_2} \cdots \phi^{i_q} \right)} \\
J\text{-type model, } & \quad \sqrt{2} J_{i a_1 \cdots a_q}\cdot \left(q\lambda^i \psi^{a_1} \phi^{a_2} \cdots \phi^{a_q} +  G^i \phi^{a_1} \cdots \phi^{a_q}\right)
\end{align*}
The auxiliary field $B$, introduced to linearize the $|E|^2$ term in the E-type model, plays the exact same algebraic role as the intrinsic auxiliary field $G$ in the J-type model. In contrast, the intrinsic auxiliary field $G$ in the E-type model completely decouples from the disorder interactions. Consequently, when establishing a duality between the two models, the mapping must explicitly identify the introduced field $B$ in the E-type model with the intrinsic field $G$ in the J-type model.

First, we must carefully handle the Gaussian integration over the complex coefficients $J_{ia_1\cdots a_q}$. Using the integral formula $\langle e^{(-J X - \bar{J} Y)} \rangle_{(0,\sigma^2)}\propto e^{ \left(\sigma^2 (X Y+YX\right)}$, we obtain the interaction part of the effective action after ensemble averaging over the disorder and have
\begin{align*}
S_{\text{int}} = & \iint d^2z d^2z' \sum_{i} \bar{\lambda} (\psi_{i_1} \phi_{i_2} \ldots \phi_{i_q} + \ldots + \phi_{i_1} \ldots \phi_{i_{q-1}} \psi_{i_q})\big|_{z'} \nonumber \\
& \times (\bar{\psi}_{i_q}\bar{\phi}_{i_{q-1}} \ldots \bar{\phi}_{i_1} + \ldots + \bar{\phi}_{i_q}\bar{\phi}_{i_{q-1}} \ldots \bar{\psi}_{i_1})\lambda \big|_{z} \\
& + \iint d^2z d^2z' \sum_{i} (\bar{\psi}_{i_q}\bar{\phi}_{i_{q-1}} \ldots \bar{\phi}_{i_1} + \ldots + \bar{\phi}_{i_q}\bar{\phi}_{i_{q-1}} \ldots \bar{\psi}_{i_1})\lambda \big|_{z} \nonumber \\
& \times \bar{\lambda} (\psi_{i_1} \phi_{i_2} \ldots \phi_{i_q} + \ldots + \phi_{i_1} \ldots \phi_{i_{q-1}} \psi_{i_q})\big|_{z'}. 
\end{align*}

To express the effective action in terms of bi-local fields, we introduce the macroscopic Green's functions $G^\Psi(z,z')$ and the self-energies $\Sigma^\Psi(z,z')$ as Lagrange multipliers. This is rigorously achieved by inserting the following identity into the path integral for each field component $\Psi \in \{\phi, \psi, \lambda, B\}$:
\begin{equation*}
    1 \propto \int \mathcal{D}\Sigma \mathcal{D}G \exp\left( - \iint d^2z d^2z' \sum_{\Psi} \Sigma^\Psi(z,z') \left[ N_\Psi G^\Psi(z,z') - \sum_{a=1}^{N_\Psi} \bar{\Psi}^a(z)\Psi^a(z') \right] \right),
\end{equation*}
where the flavor number $N_\Psi$ is $N$ for the components of chiral multiplets ($\phi, \psi$) and $M$ for the components of Fermi multiplets ($\lambda, B$). This integral identity dynamically enforces the definition $G^\Psi(z,z') = \sum_a \langle \bar{\Psi}^a(z)\Psi^a(z') \rangle/N_\Psi $. By applying this relation, we can formally replace the microscopic field bilinears in the disorder-averaged interaction with the macroscopic bi-local fields. Special care must be taken with the contraction of fermionic fields, such as $\lambda$ and $\bar{\lambda}$—particularly regarding the ordering of spatial arguments $z$ and $z'$, as well as the signs associated with the anti-commutativity of Grassmann variables. After performing the ensemble average over the random couplings, the interaction part of the resulting effective action is thus  rewritten as:

\begin{align}\label{eq:S_eff}
S_{G\Sigma} = &-\iint d^2z d^2z' N \bigg(\Sigma^{\psi} \left(G^{\psi}-\frac{1}{N}\sum \bar\psi\psi\right)+ \Sigma^{\phi}\left(G^{\phi}-\frac{1}{N}\sum \bar\phi\phi \right)\bigg) \nonumber \\
&-\iint d^2z d^2z' M \bigg(\Sigma^{\bar\lambda} \left(G^{\bar\lambda}-\frac{1}{M}\sum \lambda\bar\lambda\right)+ \Sigma^{B}\left(G^{B}-\frac{1}{M}\sum \bar BB \right)\bigg) \nonumber \\
&+\int d^2z \left(4 \bar{\phi} \partial^2 \phi -2 \bar{\psi} \partial \psi - 2 \bar{\lambda} \partial \lambda + \bar{G} G + \bar{B}B\right) \nonumber \\
&+\iint d^2z d^2z' \, M J^2 G^{\bar\lambda}(z,z') G^{\psi}(z,z') (G^{\phi}(z,z'))^{q-1} \nonumber \\
& +\iint d^2z d^2z' \, \frac{J^2M}{q} G^{B}(z,z') (G^{\phi}(z,z'))^q.
\end{align}

\subsection{Two-Point Functions}

 We focus on the low-energy regime $\omega \ll 1 \ll J$, where kinetic terms can be neglected. By taking the functional variation of this action with respect to the bi-local fields $G^\Psi(z_1,z_2)$ and self-energies $\Sigma^\Psi(z_1,z_2)$, we obtain the Schwinger-Dyson (SD) equations.
\begin{align}\label{EtypeSD}
	&\int d^2 z_2\Sigma^{\Psi}(z_1,z_2)G^{\Psi}(z_2,z_3)=-\delta(z_1,z_3)\quad \Psi\in (\phi,\psi,\lambda,B,G),\\
	&\Sigma^{\psi}(z_1,z_2) = \mu J^2 G^{\bar{\lambda}}(z_1,z_2) (G^{\phi}(z_1,z_2))^{q-1}, \\
	&\Sigma^{\phi}(z_1,z_2) = J^2 \mu \bigg( (q-1) G^{\bar{\lambda}}(z_1,z_2) G^{\psi}(z_1,z_2) (G^{\phi}(z_1,z_2))^{q-2} \nonumber \\
	&\qquad\qquad\qquad\qquad + G^B(z_1,z_2) \left( G^{\phi}(z_1,z_2) \right)^{q-1} \bigg),  \\
	&\Sigma^{\bar{\lambda}}(z_1,z_2) = J^2 G^{\psi}(z_1,z_2) (G^{\phi}(z_1,z_2))^{q-1}, \\
	&\Sigma^B(z_1,z_2) = \frac{J^2}{q} (G^\phi(z_1,z_2))^q,\\
	&\Sigma^G(z_1,z_2) =0.
\end{align}
 The intrinsic auxiliary field $G$ lacks interaction couplings, leading to a trivial zero self-energy ($\Sigma^G = 0$). Observing the homogeneity of the SD equations, we will adopt a conformal ansatz,
\begin{equation}
G^i(z_1, z_2) = \frac{n_i}{(z_1 - z_2)^{2h_i}(\bar{z}_1 - \bar{z}_2)^{2\tilde{h}_i}}.
\end{equation}

In the conformal regime, the bi-local fields depend only on the coordinate difference. We will use the notation $z_{12} \equiv z_1 - z_2$ in our subsequent discussions. For the fermionic fields, Grassmann anti-commutativity dictates that $G^{\bar{\lambda}}(z_{12}) =\frac{1}{M} \sum_a\langle \lambda(z_1)^a \bar{\lambda}(z_2)^a \rangle = - \frac{1}{M}\sum_a\langle  \bar{\lambda}^a (z_2) \lambda^a(z_1) \rangle = -G^{\lambda}(-z_{12})$. Meanwhile, the conformal ansatz requires $G^{\lambda}(-z_{12}) = (-1)^{2h_\lambda-2\bar h_\lambda} G^{\lambda}(z_{12})$. Since the fermion carries a half-integer spin $s = h - \bar{h}$, we have $(-1)^{2h_\lambda-2\bar h_\lambda}=-1$. These two negative signs perfectly cancel out, yielding the strict identity $G^{\bar{\lambda}}(z_{12}) = G^{\lambda}(z_{12})$. This allows us to safely replace $G^{\bar{\lambda}}$ with $G^{\lambda}$ (and correspondingly $\Sigma^{\bar{\lambda}}$ with $\Sigma^{\lambda}$) in the IR SD equations.

Nevertheless, we observe the structure is greatly similar to $J$-type model. It not only yields identical conformal weights $(h, \tilde{h})$ for the $\phi, \psi, \lambda$ fields,  auxiliary field $B$ also exhibits behavior highly consistent with the field $G$ in the $J$-type model. Specifically, the conformal weight and prefactor $n_B$ are identical to those of $G$ and detailed calculations are provided in Appendix \ref{app}.
\begin{equation}
	\begin{aligned}
		h_\phi &= \frac{\mu q - 1}{2\mu q^2 - 2}, & h_\psi &= \frac{\mu q^2 + \mu q - 2}{2\mu q^2 - 2}, & h_\lambda &= \frac{q - 1}{2\mu q^2 - 2}, & h_B &= \frac{\mu q^2 + q - 2}{2\mu q^2 - 2} \\[8pt]
		\bar{h}_\phi &= \frac{\mu q - 1}{2\mu q^2 - 2}, & \bar{h}_\psi &= \frac{\mu q - 1}{2\mu q^2 - 2}, & \bar{h}_\lambda &= \frac{\mu q^2 + q - 2}{2\mu q^2 - 2}, & \bar{h}_B &= \frac{\mu q^2 + q - 2}{2\mu q^2 - 2}.
	\end{aligned}
\end{equation}

 However, since the field $B$ does not possess manifest supersymmetry, we obtain two distinct consistency relations for the prefactors $n_i$ instead of the single relation found in \cite{Peng_2018},
\begin{equation}
	\begin{aligned}\label{eq:fac_relation}
		n_{\lambda}n_{\phi}^q &= -\frac{(q-1) q}{2\pi ^2 J^2 \left(\mu q^2-1\right)}, \\
		n_B n_{\phi}^q &= \frac{(q-1)^2 q}{\pi^2 J^2 (\mu q^2 - 1)^2}.
	\end{aligned}
\end{equation}

Consequently, we conclude that within the IR regime, the Lagrangian and SD equations establish the equivalence summarized in Table \ref{model_mapping}.

\begin{table}[htbp]
    \centering
    \caption{Correspondence map between the $J$-type and $E$-type models.}
    \label{model_mapping}
    \renewcommand{\arraystretch}{1.8}
    \begin{tabular}{@{}l c c@{}}   
        \toprule
        \textbf{Type} & \textbf{$J$-type Model} & \textbf{$E$-type Model} \\
        \midrule
        \textbf{Coupling} & $J_{i a_1 \cdots a_q}$ & $\displaystyle \frac{-J_{i a_1 \cdots a_q}}{\sqrt{2}}$ \\
        \textbf{Field}    & $G$                    & $B$ \\
        \bottomrule
    \end{tabular}
\end{table}

\section{Ladder Diagrams and Kernels}
\label{sec:ladder_kernels} 

As discussed in~\cite{Kitaev:2015,Maldacena_2016,Kitaev_2018}, the four-point function in the large $N$ limit is dominated by ladder diagrams. Due to the iterative structure of these diagrams, we can define an integral kernel $K(z_1, z_2; z, z')$ such that the recursion relation is given by:

\begin{align}
	\mathcal{F}_{n+1}(z_1, z_2, z_3, z_4) = \int dz dz' \, K(z_1, z_2; z, z') \mathcal{F}_n(z, z', z_3, z_4)
\end{align}

The full four-point function is obtained by summing over all ladder diagrams, which forms a geometric series:

\begin{align}
	\mathcal{F} = \sum_{n=0}^{\infty} \mathcal{F}_n = \sum_{n=0}^{\infty} K^n \mathcal{F}_0 = \frac{1}{1-K} \mathcal{F}_0.
\end{align}

In our model, since the integral kernel mediates transitions between different fields, $K$ naturally carries explicit field indices and should be treated as a matrix $K^{ij}$.
\begin{figure}[htbp]
	\begin{center}
		\includegraphics[width=0.4\textwidth]{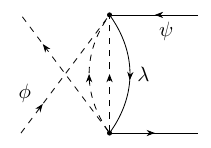}
		\includegraphics[width=0.4\textwidth]{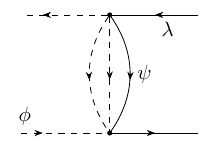}
		\includegraphics[width=0.4\textwidth]{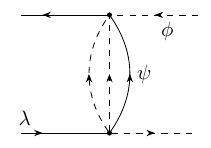}
		\caption{Diagrammatic representation of the $K^{\phi \lambda}, K^{\phi\lambda}, K^{\lambda \phi}$ kernel induced by the E-type interaction. Because the interaction vertex is given by $J_{a i_1 \cdots i_q} \bar{\lambda}_a \psi_{i_1} \phi_{i_2} \cdots \phi_{i_q} + \text{h.c.}$, the flow of the propagator arrows is distinct from that in the J-type model. This figure depicts three different kernel topologies; notably, for kernels where the external legs contain the $\lambda$ field, the ladder rails remain parallel.}
		\label{fig:E-kernel}
	\end{center}
\end{figure} 

Following the diagrammatic representation in Fig.~\ref{fig:E-kernel}, the E-type four-point functions are governed by three distinct ladder kernel topologies. To elucidate the underlying microscopic mechanics, we first focus on the detailed derivation of the $K^{\phi \psi}$ kernel. By applying Wick's theorem to the E-type interaction vertices and carefully tracking the Grassmann signs during the operator contractions, the analytical expression for this specific kernel is constructed as:
\begin{align}
	K^{\phi \psi}(z_1, z_2; z_3, z_4) = (-1) (q-1)J^2\mu G^\phi(z_{14})G^\phi(z_{32})G^\lambda(z_{43})(G^\phi(z_{34}))^{q-2}
\end{align}
This expression explicitly demonstrates how the unique arrow flow of the E-type propagators determines the specific combination of bi-local fields. Specifically, the overall $(-1)$ sign arises from fermionic operator exchanges, while the factor $(q-1)$ represents the freedom of cutting one internal $G^\phi$ propagator from the self-energy $\Sigma^\psi$.

Following the procedure outlined above, we explicitly evaluate the kernels corresponding to the remaining diagrammatic topologies illustrated in Fig.~\ref{fig:E-kernel}.
\begin{align*}
	K^{\phi\phi}(z_1, z_2; z_3, z_4) &= (q-1)J^2 \mu G^\phi(z_{14}) G^\phi(z_{32}) G^B(z_{34}) (G^\phi(z_{34}))^{q-2}\\
	&\quad-(q-1)(q-2) J^2 \mu G^\phi(z_{14}) G^\phi(z_{32}) G^\psi(z_{34}) G^{\lambda}(z_{34}) (G^\phi(z_{43}))^{q-3} \\
	K^{\phi\psi}(z_1, z_2; z_3, z_4) &= -(q-1)J^2\mu G^\phi(z_{14})G^\phi(z_{32})G^\lambda(z_{43})(G^\phi(z_{34}))^{q-2} \\
	K^{\phi\lambda}(z_1, z_2; z_3, z_4) &= -(q-1)J^2 G^\phi(z_{13}) G^\phi(z_{42}) G^\psi(z_{43}) (G^\phi(z_{43}))^{q-2} \\
	K^{\phi B}(z_1, z_2; z_3, z_4) &= J^2 G^\phi(z_{14}) G^\phi(z_{32}) (G^\phi(z_{34}))^{q-1} \\
	K^{\psi\phi}(z_1, z_2; z_3, z_4) &= (q-1)J^2 \mu G^\psi(z_{14}) G^\psi(z_{32}) G^{\lambda}(z_{43}) (G^\phi(z_{34}))^{q-2} \\
	K^{\psi\lambda}(z_1, z_2; z_3, z_4) &= J^2 G^\psi(z_{13}) G^\psi(z_{42}) (G^\phi(z_{43}))^{q-1} \\
	K^{\lambda\phi}(z_1, z_2; z_3, z_4) &= (q-1)J^2 \mu G^{\lambda}(z_{13}) G^{\lambda}(z_{42}) G^\psi(z_{34}) (G^\phi(z_{34}))^{q-2} \\
	K^{\lambda\psi}(z_1, z_2; z_3, z_4) &= J^2 \mu G^{\lambda}(z_{13}) G^{\lambda}(z_{42}) (G^\phi(z_{34}))^{q-1} \\
	K^{B\phi}(z_1, z_2; z_3, z_4) &= J^2 \mu G^B(z_{14}) G^B(z_{32}) (G^\phi(z_{34}))^{q-1}
\end{align*}

Adopting the following ansatz for the eigenfunctions,
\begin{align*}
	\Phi^i(z_1, z_2) &= (z_{12})^{h-2h_i}(\bar{z}_{12})^{\tilde{h}-2\tilde{h}_i}, \quad i \in \{\phi, \psi, \lambda, B\},
\end{align*}
we evaluate the kernel matrix elements $k^{ij}$ through the convolution equation$$K^{(ij)} * \Phi^j = k^{ij} \Phi^i.$$
In evaluating the kernel matrix elements via convolution, we frequently encounter integrals of the following form:
\begin{align}
	I_{total} = \int d^2x_3 d^2x_4 (x_{14})^{-2h_i}(\bar{x}_{14})^{-2\bar{h}_i} (x_{23})^{-2h_i}(\bar{x}_{23})^{-2\bar{h}_i} (x_{43})^{h - 2 + 2h_i}(\bar{x}_{43})^{\tilde{h} - 2 + 2\bar{h}_i}
\end{align}
For brevity, we state the final result of this integration directly below, deferring the detailed derivation to Appendix~\ref{app}:
\begin{equation}\label{eq:inteq2}
	\begin{aligned}
		I_{total}& = k(h,\tilde h, h_i, \bar h_i)(x_{12})^{h-2h_i}(\bar x_{12})^{\tilde h-2\bar h_i},\\
		k(h,\tilde h,h_i,\bar h_i)&\equiv \pi^2 \left[ \frac{\Gamma(2h_i - h)}{\Gamma(2h_i)^2 \Gamma(2 - h - 2h_i)} \right] \left[ \frac{\Gamma(1 - 2\bar{h}_i)^2 \Gamma(\tilde{h} - 1 + 2\bar{h}_i)}{\Gamma(\tilde{h} + 1 - 2\bar{h}_i)} \right].
	\end{aligned}
\end{equation}

To apply the integral formula in Eq.~\eqref{eq:inteq2}, we must rearrange the bi-local functions within the integrand so that their coordinate indices form a sequentially connected chain. Taking the $K^{\phi\psi}$ as an explicit example:
\begin{align}
	&\frac{1}{n_\phi^{q}n_\lambda}\int d^2z_3 d^2z_4 K^{\phi\psi}(z_1, z_2; z_3, z_4) \Phi^{\psi}(z_3,z_4)\nonumber \\
	=& (q-1)\int d^2z_3 d^2z_4 (z_{14})^{-2h_\phi} (z_{32})^{-2h_\phi} (z_{43})^{-2h_\lambda} (z_{34})^{-2(q-2)h_\phi}(z_{34})^{h-2h_\psi}\nonumber \\ & (\bar z_{14})^{-2\bar h_\phi} (\bar z_{32})^{-2\bar h_\phi} (\bar z_{43})^{-2\bar h_\lambda} (\bar z_{34})^{-2(q-2)\bar h_\phi}(z_{34})^{\tilde h-2\bar h_\psi}\nonumber \\
	=&(q-1)(-1)^{h-\tilde{h}-2(h_\psi-\bar h_\psi) -2(q-2)(h_\phi-\bar h_\phi)}\int d^2z_3 d^2z_4  (z_{14})^{-2h_\phi} (z_{32})^{-2h_\phi} \nonumber \\ &(z_{43})^{h-2h_\lambda-2(q-2)h_\phi-2h_\psi}  (\bar z_{14})^{-2\bar h_\phi} (\bar z_{32})^{-2\bar h_\phi} (\bar z_{43})^{\tilde{h}-2\bar h_\lambda-2(q-2)\bar h_\phi-2\bar h_\psi}\nonumber\\
	=&(q-1)(-1)^{h-\tilde{h}-1}k(h,\tilde{h},h_\phi,\tilde{h}_\phi)\Phi^\phi(z_1,z_2) 
\end{align}
As explicitly shown in the second equality, rewriting the coordinate difference $(z_{34})$ as $(-z_{43})$ to form a sequentially connected integration chain naturally extracts the specific phase factor $(-1)^{h-\tilde{h}-1}$. 

Following the same procedure, by carefully evaluating the specific propagator directions in the remaining Feynman diagrams and utilizing Eq.~\eqref{eq:inteq2}, we obtain the explicit expressions for all $k^{ij}$:

\begin{align}
	k^{\phi\phi} &=(-1)^{h-\tilde h} \frac{\pi^2\mu J^2(q-1)n_\phi^qn_B \Gamma(1 - 2h_\phi)^2 \Gamma(2h_\phi - h) \Gamma(\tilde{h} - 1 + 2h_\phi)}{ \Gamma(2h_\phi)^2 \Gamma(2 - 2h_\phi - h) \Gamma(\tilde{h} + 1 - 2h_\phi)}\nonumber \\[10pt]
	&\quad+(-1)^{h-\tilde h}\frac{\pi^2\mu J^2(q-1)(q-2)n_\phi^{q-1}n_\lambda n_\psi \Gamma(1 - 2h_\phi)^2 \Gamma(2h_\phi - h) \Gamma(\tilde{h} - 1 + 2h_\phi)}{ \Gamma(2h_\phi)^2 \Gamma(2 - 2h_\phi - h) \Gamma(\tilde{h} + 1 - 2h_\phi)}, \\[10pt]
	k^{\phi\psi} &=(-1)^{h-\tilde h} \frac{\pi^2\mu J^2(q-1) n_\phi^q n_\lambda \Gamma(1 - 2h_\phi)^2 \Gamma(2h_\phi - h) \Gamma(\tilde{h} - 1 + 2h_\phi)}{ \Gamma(2h_\phi)^2 \Gamma(2 - 2h_\phi - h) \Gamma(\tilde{h} + 1 - 2h_\phi)},\\[10pt]
	k^{\phi\lambda} &=\frac{\pi^2 J^2(q-1)n_\phi^{q} n_\psi\Gamma(1 - 2h_\phi)^2 \Gamma(2h_\phi - h) \Gamma(\tilde{h} - 1 + 2h_\phi)}{\Gamma(2h_\phi)^2 \Gamma(2 - 2h_\phi - h) \Gamma(\tilde{h} + 1 - 2h_\phi)},  \\[10pt]
	k^{\phi B} &= (-1)^{h-\tilde h}\frac{\pi^2 J^2 n_\phi^{q+1}\Gamma(1 - 2h_\phi)^2 \Gamma(2h_\phi - h) \Gamma(\tilde{h} - 1 + 2h_\phi)}{\Gamma(2h_\phi)^2 \Gamma(2 - 2h_\phi - h) \Gamma(\tilde{h} + 1 - 2h_\phi)},  \\[10pt]
	k^{\psi\phi} &= (-1)^{h-\tilde h}\frac{(q-1)\pi^2\mu J^2n_\psi^2 n_\lambda n_\phi^{q-2} \Gamma(1 - 2h_\phi)^2 \Gamma(1 + 2h_\phi - h) \Gamma(\tilde{h} - 1 + 2h_\phi)}{\Gamma(1 + 2h_\phi)^2 \Gamma(1 - 2h_\phi - h) \Gamma(\tilde{h} + 1 - 2h_\phi)},  
\end{align}
\begin{align}
	k^{\psi\lambda} &= \frac{\pi^2 J^2n_\psi^2 n_\phi^{q-1}\Gamma(1 - 2h_\phi)^2 \Gamma(1 + 2h_\phi - h) \Gamma(\tilde{h} - 1 + 2h_\phi)}{ \Gamma(1 + 2h_\phi)^2 \Gamma(1 - 2h_\phi - h) \Gamma(\tilde{h} + 1 - 2h_\phi)},  \\[10pt]
	k^{\lambda\phi} &= \frac{(q-1)\pi^2 \mu J^2  n_\phi^{q-2}n_\lambda^2 n_\psi\Gamma(-2h_\lambda)^2 \Gamma(2h_\lambda - h) \Gamma(2h_\lambda + \tilde{h})}{ \Gamma(2h_\lambda)^2 \Gamma(2 - h - 2h_\lambda) \Gamma(\tilde{h} - 2h_\lambda)},  \\[10pt]
	k^{\lambda\psi} &= \frac{\pi^2\mu J^2 n_\phi^{q-1} n_\lambda^2 \Gamma(-2h_\lambda)^2 \Gamma(2h_\lambda - h) \Gamma(2h_\lambda + \tilde{h})}{ \Gamma(2h_\lambda)^2 \Gamma(2 - h - 2h_\lambda) \Gamma(\tilde{h} - 2h_\lambda)},\\
	k^{B\phi} &=(-1)^{h-\tilde h} \frac{\pi^2\mu J^2 n_\phi^{q-1}n_B^2 \Gamma(-2h_\lambda)^2 \Gamma(1-h+2h_\lambda) \Gamma(2h_\lambda + \tilde{h})}{ \Gamma(1+2h_\lambda)^2 \Gamma(1-h-2h_\lambda) \Gamma(\tilde{h} - 2h_\lambda)} . 
\end{align}

Incorporating the disorder coupling contributions and the factors $n_i$, we construct the full Kernel Matrix for the E-type model. However, a naive element-by-element comparison with the J-type kernel matrix is inadequate. First, one cannot assume that the exact value of the prefactor $n_i$ is trivially identical across the two distinct ensembles. Second, the distribution of $(-1)^{h-\tilde h}$ phases differs significantly: owing to its specific interaction vertex, every matrix element $k^{ij}$ in the J-type model universally acquires a phase factor of $(-1)^{h-\bar{h}}$. In contrast, the E-type kernel only exhibits this $(-1)^{h-\bar{h}}$ phase in components where neither the incoming nor the outgoing field index involves $\lambda$.

Therefore, to rigorously establish the dynamical equivalence between the two theories, we must shift our focus to the characteristic roots, which are obtained by solving the equation $E(x, h, \bar{h}, \mu, q) = 0$. Here, the characteristic polynomial is explicitly given by:
\begin{equation}
	\begin{split}
		E(x, h, \bar{h}, \mu, q) &\equiv \det\bigg(k^{ij}(h,\bar{h},\mu,q)-x\cdot \mathds{1}\bigg) \\
		&= x^4 - k^{\phi\phi}x^3 - \left( k^{\phi B}k^{B\phi} + k^{\phi\psi}k^{\psi\phi} + k^{\phi\lambda}k^{\lambda\phi} + k^{\psi\lambda}k^{\lambda\psi} \right) x^2 \\
		&\quad + \left( k^{\phi\phi}k^{\psi\lambda}k^{\lambda\psi} - k^{\phi\psi}k^{\psi\lambda}k^{\lambda\phi} - k^{\phi\lambda}k^{\psi\phi}k^{\lambda\psi} \right) x + k^{\phi B}k^{\psi\lambda}k^{\lambda\psi}k^{B\phi} \, .
	\end{split}
\end{equation}

Utilizing the algebraic consistency relations derived in Eq.~\eqref{eq:fac_relation}, we can demonstrate that the characteristic roots of both models are entirely independent of the specific coupling $J$ and the prefactor $n_\phi$. Remarkably, despite the distinct internal phase structures of their respective kernels, their characteristic determinants are strictly identical. This profound mathematical property guarantees that the E-type model preserves the higher-spin symmetry, much like the J-type model, a feature we will explore extensively in the subsequent chapter.

%% file: Main/J_type_an_higher_spin.tex
\section{Higher-Spin Symmetry in the $E$-Type Model}\label{Jtype}

In this section, having established the consistency of the characteristic determinants, we briefly sketch how to employ the kernel matrix to probe the properties of higher spin symmetry in the IR region. For a detailed discussion, please refer to~\cite{Peng_2018}.

\subsection{Ladder Equation and the Pole Condition}

As we said before, the full four-point function is obtained by summing over all ladder diagrams, which forms a geometric series:

$$
\mathcal{F} = \sum_{n=0}^{\infty} \mathcal{F}_n = \sum_{n=0}^{\infty} K^n \mathcal{F}_0 = \frac{1}{1-K} \mathcal{F}_0.
$$

At the same time, the four-point function can be expressed via the Operator Product Expansion (OPE) as a sum over primary operators. Consequently, the pole condition $k=1$ identifies the physical primary operators propagating between two channels of the ladder diagram.

Since we have already absorbed the phase factor $(-1)^{h-\tilde{h}}$ into $k^{ij}$, there is no need to distinguish between symmetric and antisymmetric channels as in Ref.~\cite{Peng_2018}. Therefore, studying the operators with specific $(h,\tilde{h})$ propagating in the channels is equivalent to solving:
\begin{equation*}
	E(x=1, h, \tilde{h}, \mu, q) = 0 .
\end{equation*}

\subsection{Evidence of Higher Spin Symmetry}

To systematically investigate the emergence of higher-spin symmetry in this model, we recall that a conserved holomorphic operator of spin $s$ is characterized by the conformal weights $(h, \tilde{h}) = (s, 0)$. Therefore, we can fix the spin $h - \tilde{h} = s$ (i.e., $h = s + \tilde{h}$) and numerically solve the characteristic equation $E(x=1, s+\tilde{h}, \tilde{h}, \mu, q) = 0$ to track the trajectory of $\tilde{h}$ as a function of the parameter $\mu$. If $\tilde{h} \rightarrow 0$ as $\mu$ approaches the critical limits, it serves as a definitive signature for the existence of a conserved holomorphic current. Similarly, for an anti-holomorphic operator of spin $s$, we set $\tilde{h} = s + h$ and solve $E(x=1, h, s+h, \mu, q) = 0$ to verify whether $h \rightarrow 0$. 

\begin{figure}[htbp]
	\centering
	
	\begin{subfigure}[b]{0.48\textwidth}
		\centering
		\includegraphics[width=\textwidth]{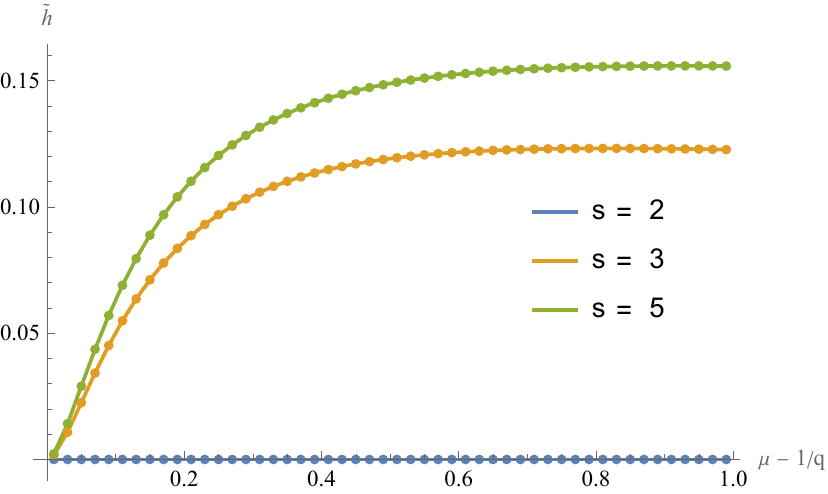} 
		\caption{Holomorphic, $\mu \to (1/q)^+$}
		\label{fig:holo_1_over_q}
	\end{subfigure}
	\hfill
	\begin{subfigure}[b]{0.48\textwidth}
		\centering
		\includegraphics[width=\textwidth]{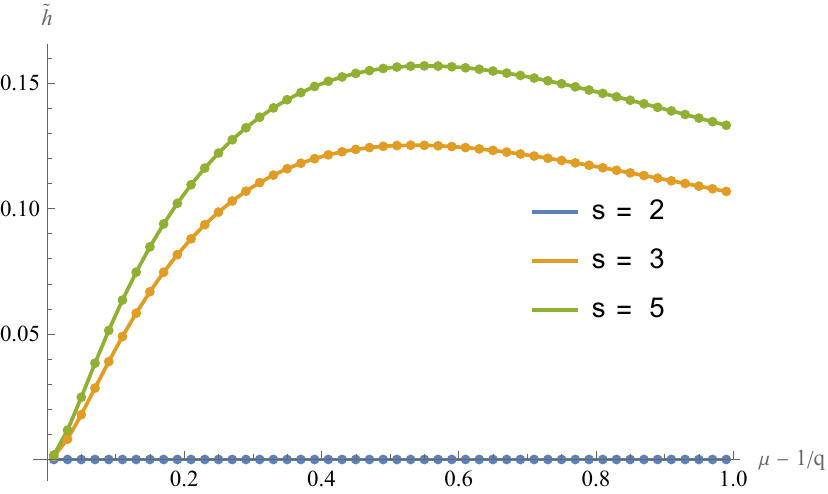} 
		\caption{Anti-holomorphic, $\mu \to (1/q)^+$}
		\label{fig:anti_holo_1_over_q}
	\end{subfigure}
	
	\vspace{0.5cm}

	\begin{subfigure}[b]{0.48\textwidth}
		\centering
		\includegraphics[width=\textwidth]{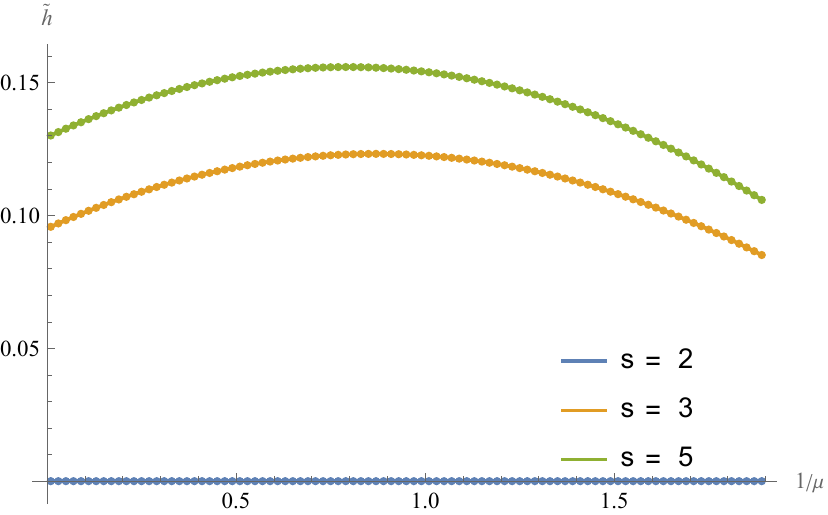} 
		\caption{Holomorphic, $\mu \to +\infty$}
		\label{fig:holo_infty}
	\end{subfigure}
	\hfill
	\begin{subfigure}[b]{0.48\textwidth}
		\centering
		\includegraphics[width=\textwidth]{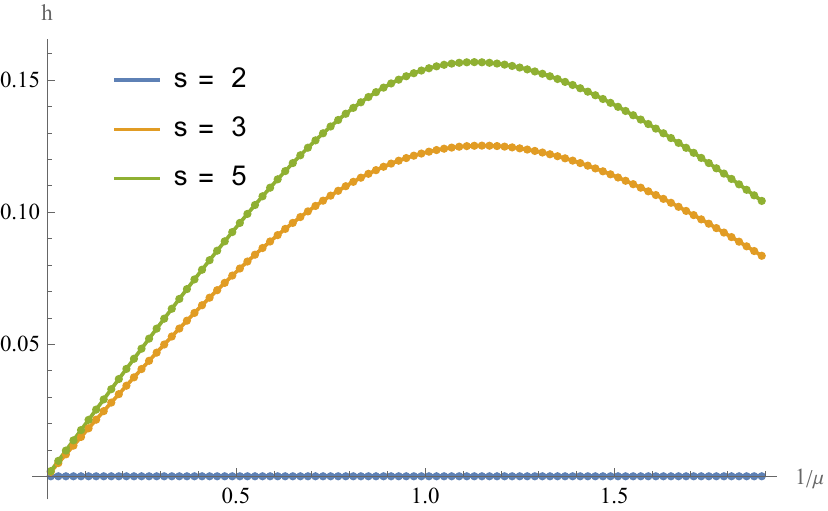}
		\caption{Anti-holomorphic, $\mu \to +\infty$}
		\label{fig:anti_holo_infty}
	\end{subfigure}
	
	\caption{The continuous evolution of the anomalous dimensions for various spin operators as a function of $\mu$ at fixed $q=3$. The top row (panels a and b) illustrates the behavior as $\mu$ approaches the critical limit $(1/q)^+$, while the bottom row (panels c and d) displays the asymptotic behavior as $\mu \to +\infty$. With the explicit exception of the holomorphic operators with spins $s=3$ and $s=5$ in the $\mu \to +\infty$ limit (which are not expected to be conserved for $q=3$), the anomalous dimensions exactly vanish in all other respective cases. This visually confirms the selective emergence of the higher-spin conserved currents exactly as summarized in Table~\ref{tab:spin_distribution}.}
	\label{fig:anomalous_dim}
\end{figure}

Following the numerical methodology established in Ref.~\cite{Peng_2018}, our verification explicitly confirms that in the limit $\mu\rightarrow (1/q)^+$, both $\tilde{h}$ and $h$ indeed vanish, yielding $(s, 0)$ and $(0, s)$ as exact solutions for integer spins $s \in \mathds{Z}^+$. This property directly indicates that a robust tower of conserved higher-spin currents emerges. Similar higher-spin phenomena emerge in the $\mu\rightarrow \infty$ limit, though subject to stricter constraints on $q$. To provide a clear overview, we summarize the explicit spin distributions of these conserved operators under both limits in Table~\ref{tab:spin_distribution}. Furthermore, the continuous evolution of the anomalous dimensions for different spin operators as a function of $\mu$ is visually illustrated in Fig.~\ref{fig:anomalous_dim}. Crucially, these numerical results for the $E$-type model perfectly match those obtained for the $J$-type model. This agreement provides new evidence for the duality between the two theories.
\begin{table}[htbp]
	\centering
	\caption{Spin distributions of conserved operators under different limits.}
	\label{tab:spin_distribution}
	\begin{tabular}{|l|c|c|}
		\hline
		& holomorphic operators & anti-holomorphic operators \\
		\hline
		$\mu \to (1/q)^+$ & $(h,\tilde{h})=(s,0), s\ge 1$ & $(h,\tilde{h})=(0,s), s\ge 1$ \\
		\hline
		$\mu \to +\infty$ & $(h,\tilde{h})=(s,0), s> 2, q=2$ or $s=1,2$ & $(h,\tilde{h})=(0,s), s\ge 1$ \\
		\hline
	\end{tabular}
\end{table}

Beyond identifying the physical spectrum, the characteristic equation encodes profound dynamical information. For instance, the Lyapunov exponent $\lambda_L$, calculated via analytic continuation to the out-of-time-ordered correlator (OTOC), identically vanishes in these limits, serving as additional evidence for the restoration of symmetry. Significantly, perturbing $\mu$ away from the critical values enables the calculation of anomalous dimensions. In the large-spin limit, this yields a dispersion relation highly consistent with that of classical rotating strings in AdS spacetime, thereby reinforcing the holographic gravitational duality.

Since the characteristic determinant of the $E$-type model is strictly identical to that of the $J$-type model, detailed analyses of these features can be seamlessly mapped over. Therefore, for extensive discussions regarding the asymptotic behavior of $\mu$, the regularization of divergences near $1/q$, and the underlying connections to higher-spin $\mathcal{W}$-algebras, we refer the reader to Ref.~\cite{Peng_2018}.

%% file: Main/summary.tex
\section{Conclusion and Outlook}\label{summary}

In this work, we investigate the $E$-type configuration of the disordered model, establishing a structural duality with the $J$-type model. By employing the $(0,2)$ Landau-Ginzburg theory alongside a rigorous analysis of the Schwinger-Dyson equations—specifically through the behavior of Green functions in the conformal regime—we demonstrate that the kernel matrix's characteristic determinant remains independent of the coupling.Furthermore, by explicitly constructing the integral kernels from Feynman diagrams, we numerically extract the anomalous dimensions of the operators. The results confirm that these dimensions exactly vanish in the asymptotic limits, which verifies the emergent higher-spin symmetry and the characteristic vanishing chaos of the $E$-type model in the large-$N$ limit.

An important element of our analysis is the exchange symmetry between the fields $\lambda$ and $\bar{\lambda}$. Although our derivation is based on the Lagrangian formalism, this feature resonates with the $E \leftrightarrow E^\vee$ replacement observed in $(0,2)$ A/B-type twists \cite{Katz:2004nn}. Such a correspondence suggests that the $\lambda \leftrightarrow \bar{\lambda}$ symmetry may have a deeper geometric origin, potentially linked to Hori-Vafa-type mirror dualities for $(0,2)$ theories. While a formal geometric mapping remains beyond the scope of this paper, it hints at a broader $(0,2)$ mirror framework underpinning these disordered systems.

By extending the moduli space, our results provide another potential piece of evidence suggesting a holographic duality between SYK-like models and tensionless string theory. Looking forward, our explicit derivation of the pure $E$-model provides the essential mathematical basis for investigating the more generic mixed $E-J$ model. A highly fascinating direction is to explore how the non-trivial combinations of Feynman diagrams in this mixed system affect the physical spectrum, specifically how the number of conserved higher-spin operators evolves as one interpolates across the parameter space between the pure $E$ and pure $J$ limits. Additionally, future directions include gauging the $U(1)$ symmetry of the model as in~\cite{Zhang:2025kty}, constructing the precise bulk dual in higher-spin supergravity—specifically by establishing a connection to $\mathcal{N}=(0,2)$ higher-spin AdS$_3$ gravity via the super-Schwarzian action—and investigating how finite-$N$ corrections might disrupt the higher-spin operator spectrum.

\section*{Acknowledgements}
We are deeply grateful to Cheng Peng for suggesting this topic and useful discussions and guidance in the project. We also would like to express our sincere thanks to Eric Sharpe for his insightful comments and helpful suggestions. This work is supported by NSFC NO. 12175237, and NSFC NO. 12447108, the Fundamental Research Funds for the Central Universities, and funds from the Chinese Academy of Sciences.

%% file: Main/appendix.tex
\section{Appendix}\label{app}

\subsection*{More On Schwinger Dyson Equations}

We provide an \textit{a posteriori} justification that we can replace $G^{\bar\lambda}(z)$ with $G^{\lambda}(z)$ in the Schwinger-Dyson equations in Eq.~\eqref{EtypeSD},
\begin{align*}
\Sigma^{\psi} &= \mu J^2 G^{{\lambda}} (G^{\phi})^{q-1} \,, \\
\Sigma^{\phi} &= J^2 \mu \left[(q-1) G^{{\lambda}} G^{\psi} (G^{\phi})^{q-2} + G^B \left( G^{\phi} \right)^{q-1}\right] \,,  \\
\Sigma^{{\lambda}} &= J^2 G^{\psi} (G^{\phi})^{q-1} \,,  \\
\Sigma^B &= \frac{J^2}{q} (G^\phi)^q \,.
\end{align*}
Another set of SD equations in Fourier space is given by 

\begin{equation}\label{SD1}
    \Sigma(\omega)G(\omega)=-1 
\end{equation}

We utilize the following Fourier transform $\mathcal{F}(\cdot)$:
\begin{equation*}
\frac{1}{z^{2h}\bar{z}^{2\bar h}} \xrightarrow{\mathcal{F}} \frac{\pi}{i^{2 (h-\bar h)} 2^{2 \bar h+2 h-2} } \frac{\Gamma (1-2 h)}{\Gamma (2 \bar h)}\frac{1}{p^{2\bar h+1}\bar p^{2h+1}} \,.
\end{equation*}

According to the second set of SD equations, $\Sigma$ takes the form $\frac{n_{\small\Sigma}}{ z^{-2h_{\small\Sigma}}\bar z^{-2\tilde h_{\small\Sigma}}}$ (or linear combinations). Matching the exponents of $p$ on the RHS of Eq.~\eqref{SD1} yields the relations $h_{\Sigma}+h_{G}=1$ and $\tilde h_{\Sigma}+\tilde h_{G}=1$. Considering the spin constraint $2(h-\bar h)=2s\in \mathbb{Z}$, we proceed to solve the Fourier transformed Eq.~\eqref{SD1}:

\begin{equation*}
\mathcal{F}\left[\frac{n_{\Sigma}}{z^{2h_{\Sigma}}\bar z^{2\tilde h_{\Sigma}}}\right]\mathcal{F}\left[\frac{n_{G}}{z^{2h_{G}}\bar z^{2\tilde h_{G}}}\right] =
(-1)^{2h_{G}-2\tilde h_{G}+1}\frac{n_{\Sigma}n_{G}\pi^{2}}{(2h_{G}-1)(2\tilde{h}_{G}-1)}=-1 \,.
\end{equation*}

Applying this to the SD equations in Eq.~\eqref{EtypeSD}, we obtain the conformal weight relations (and Hermitian conjugate part),

\begin{equation}\label{hweight}
h_{\psi}+ h_{\lambda}+(q-1)h_{\phi}=1 \,,\quad h_B+qh_{\phi}=1 \,.
\end{equation}

As $h_{\lambda}$ appears independently, replacing it with $h_{\bar{\lambda}}$ preserves the equality. This confirms our earlier claim, and we can thus verify that $G^i(z)=G^{\bar{i}}(z)$ via

\begin{equation*}
\frac{n_{\bar{\Psi}}}{z^{2h_{\bar{\Psi}}}\bar{z}^{2\tilde{h}_{\bar{\Psi}}}} = (-1)^{2(h_{\Psi}-\tilde{h}_{\Psi})} \frac{n_{\Psi}}{(-z)^{2h_{\Psi}}(\overline{-z})^{2\tilde{h}_{\Psi}}}=\frac{n_{{\Psi}}}{z^{2h_{{\Psi}}}{z}^{2\tilde{h}_{{\Psi}}}} \,.
\end{equation*}

Finally, examining the coefficients $n$ in Eq.~\eqref{EtypeSD}, we derive the following coupled equations:
\begin{align}\label{eq:1}
	\frac{(-1)^{2h_{\psi}-2\tilde h_{\psi}+1} \pi^2 J^2 \mu \, n_{\psi}n_{\lambda}n_{\phi}^{q-1} }{(2h_{\psi}-1)(2\tilde{h}_{\psi}-1)} &= -1 \,,  
\end{align}
\begin{align}\label{eq:2}
	\frac{(-1)^{2h_{\lambda}-2\tilde h_{\lambda}+1} \pi^2 J^2 \, n_{\psi}n_{\lambda}n_{\phi}^{q-1} }{(2h_{\lambda}-1)(2\tilde{h}_{\lambda}-1)} &= -1 \,. 
\end{align}
Given $h_{\phi}=\tilde{h}_{\phi}$, we first use Eq.~\eqref{hweight} to express all conformal weights $(h_i,\tilde{h}_i)$ entirely in terms of $h_{\phi}$. By taking the ratio of Eq.~\eqref{eq:1} to Eq.~\eqref{eq:2}, we arrive at a constraint equation solely for $h_{\phi}$, which yields the exact solution $h_{\phi} = \frac{\mu q - 1}{2\mu q^2 - 2}$. This conformal dimension is precisely identical to that of the $J$-type model~\cite{Peng_2018}. Subsequently, by utilizing the supersymmetry relation $G_{\psi}(z_1,z_2) = -2\partial_{z_1} G_{\phi}(z_1,z_2)$, we can uniquely determine the normalization constants, yielding:
\begin{equation}
n_{\lambda}n_{\phi}^q = -\frac{q(q-1)}{2\pi ^2 J^2 \left(\mu q^2-1\right)} \,, \quad
n_B n_{\phi}^q = \frac{q(q-1)^2}{\pi^2 J^2 (\mu q^2 - 1)^2} \,.
\end{equation}

\subsection*{More On Kernel Calculation}

We consider the eigenvalue problem defined by the integral equation
\begin{equation*}
    \iint d^2 z_3 d^2 z_4 \, K^{(ij)}(z_1,z_2,z_3,z_4)\Phi^j(z_3,z_4) = k^{ij}\Phi^i(z_1,z_2),
\end{equation*}

From the explicit structure of $K^{(ij)}$, and denoting by $(h^*, \tilde{h}^*)$ the accumulated conformal weights of the Green functions connecting the two rails, the convolution with $\Phi^j$ takes the form of integration on
\begin{equation*}
    K^{(ij)} * \Phi^j \propto z_{13}^{-2h_i}\bar{z}_{13}^{-2\tilde{h}_i} z_{24}^{-2h_i}\bar{z}_{24}^{-2\tilde{h}_i} z_{34}^{h-2h_j-2h^*}\bar{z}_{34}^{\tilde{h}-2\tilde{h}_j-2\tilde{h}^*}.
\end{equation*}

By making use of the dimension relation in Eq.~\eqref{hweight}, the exponents of $z_{34}$ and $\bar{z}_{34}$ can be elegantly simplified to $h-2(1-h_i)$ and $\tilde{h}-2(1-\tilde{h}_i)$, respectively. To evaluate the integrals, we employ the standard complex integral identity:
\begin{equation}\label{eq:integral}
\begin{split}
    \int d^2y \, (y-t_0)^{a+n}&(\bar{y}-\bar{t}_0)^a (t_1-y)^{b+m}(\bar{t}_1-\bar{y})^b \\
    =& (t_1-t_0)^{a+b+n+m+1}(\bar{t}_1-\bar{t}_0)^{a+b+1} \times \\
    & \pi \frac{\Gamma(a+1)\Gamma(b+1)\Gamma(-a-b-m-n-1)}{\Gamma(a+b+2)\Gamma(-a-n)\Gamma(-b-m)}.
\end{split}
\end{equation}

By applying the integration formula given in Eq.~\eqref{eq:integral} twice, we obtain:

\begin{equation}\label{eq:inteq2}
	\begin{aligned}
		&\int d^2z_3 d^2z_4 z_{13}^{-2h_i}\bar{z}_{13}^{-2\tilde{h}_i} z_{24}^{-2h_i}\bar{z}_{24}^{-2\tilde{h}_i} z_{34}^{h-2h_j-2h^*}\bar{z}_{34}^{\tilde{h}-2\tilde{h}_j-2\tilde{h}^*}.\\ =& k(h,\tilde h, h_i, \bar h_i)(x_{12})^{h-2h_i}(\bar x_{12})^{\tilde h-2\bar h_i},\\
		k(h,\tilde h,h_i,\bar h_i)&\equiv \pi^2 \left[ \frac{\Gamma(2h_i - h)}{\Gamma(2h_i)^2 \Gamma(2 - h - 2h_i)} \right] \left[ \frac{\Gamma(1 - 2\bar{h}_i)^2 \Gamma(\tilde{h} - 1 + 2\bar{h}_i)}{\Gamma(\tilde{h} + 1 - 2\bar{h}_i)} \right],
	\end{aligned}
\end{equation}
which precisely matches the assumed eigenfunction structure $K^{ij} * \Phi^j \propto \Phi^{i}$. This confirms the consistency and correctness of the proposed eigenfunctions. This methodology can be extended to out-of-time-ordered correlators (OTOCs) by utilizing different Green's functions. In this context, specific reparametrizations are preferred to ease the computational burden; please refer to Refs.~\cite{Murugan:2017eto, Peng_2018} for detailed discussions.